\documentclass[traditabstract]{aa}
\usepackage{graphicx}
 \usepackage{lscape}
 \usepackage{colortbl}
 \usepackage{natbib}
 \usepackage{verbatim} 
  \usepackage{comment} 
\usepackage{ulem}
\usepackage{url}
 \bibpunct{(}{)}{;}{a}{}{,}
\usepackage{txfonts}
\begin{document}
\input psfig.tex
\def\simgt{\stackrel{>}{{}_\sim}}
\def\simlt{\stackrel{<}{{}_\sim}}

\titlerunning{Metal history through Bayes}

\title{The enrichment history of the intracluster medium: a Bayesian 
approach\thanks{Table 1 only available in electronic form
at the CDS}}
\author{S. Andreon} 
\institute{
INAF--Osservatorio Astronomico di Brera, via Brera 28, 20121, Milano, Italy\\
\email{stefano.andreon@brera.inaf.it} 
}
\date{Received --, 2012; accepted --, 2012}

\abstract{
This work measures the evolution of the iron content in galaxy clusters by a
rigorous analysis of the data of 130 clusters at $0.1<z<1.3$. This task
is made difficult by a) the low signal--to--noise ratio of 
abundance measurements and the upper limits,
b) possible selection effects, c) boundaries in the parameter 
space, d) non--Gaussian errors, e) the intrinsic variety of the
objects studied, and f) abundance systematics. We introduce a Bayesian model 
to address all these issues at the
same time, thus allowing cross--talk (covariance). On simulated data, the
Bayesian fit
recovers the input enrichment history, unlike in standard
analysis.
After accounting for a possible dependence on X--ray temperature, for 
metal abundance systematics, and for the intrinsic variety of
studied objects, we found that
the present--day metal content is not reached
either at high or at low redshifts, but gradually over time:
iron abundance increases by a factor 1.5 in the 7 Gyr sampled by the data.
Therefore, feedback in metal abundance does not end at high 
redshift. Evolution is established with a moderate amount of
evidence, 19 to 1 odds against
faster or slower metal enrichment histories.
We quantify, for the first time, the intrinsic spread 
in metal abundance, $18\pm3$ \%, after correcting for the effect of
evolution, X--ray temperature, and metal abundance systematics.
Finally, we also present an analytic approximation of the X--ray temperature
and metal abundance likelihood functions, which are
useful for other regression fitting
involving these parameters. The data for the 130 clusters and code used for
the stochastic computation are provided with the paper. }

   \maketitle

\section{Introduction}

The physics of galaxy clusters is complex because of the
interplay of cosmology (structure growth), 
gas physics, star formation (and its feedback), and possibly
AGN. We lack an ab initio theory able to make predictions which are
not falsified by data, about how
normal baryonic matter collects in dark matter gravitational 
potential wells and forms
galaxies, which in turn influence the intracluster medium 
(e.g. Young et al. 2011 and references therein).
For example, star formation (galaxy feedback) is a suspected
source of non--gravitational entropy excess (Buote et al. 2007; 
Pratt et al. 2006; Sun et al. 2009), yet the amount
of star formation required to reproduce 
the observed X--ray derived quantities (e.g. 
baryon fraction in gas or mass--temperature scaling relations) is at least
ten times greater than the
amount of stars observed (Gonzalez et al. 2007; Kravtsov
et al. 2009; Andreon 2010). 

Owing to the complex behavior of the baryonic matter and the
lack of a satisfactory ab initio theory, we can only
progress by 
observational studies of the evolution of the intracluster medium.
Measuring the evolution of the gas metallicity is
key information for understanding when metals are produced and
also at which time the stellar feedback enriches the intracluster medium
(e.g. Ettori 2005). Gas metallicity is usually 
inferred from the fit of the cluster X--ray spectrum, and the 
derived metal abundance is, largely, the abundance in iron, because
data of much higher quality than usually available
are needed to constrain the abundance of the other elements.

Determination of the evolution of intracluster medium
metal content (Fe, indeed) has been already
addressed by previous works, e.g. Balestra et al. (2007), Maughan et al.
(2008), Anderson et al. (2009), and Baldi et al (2011),
to cite the most recent ones. Key points to
be addressed are related to the low--quality determination of most measurements,
the possible impact of selection effects, the intrinsic variety of studied
objects, the presence of boundaries in the data and parameter space, 
and metal abundance systematics.
Therefore, a proper analysis of the data is mandatory and
should be guided by statistical considerations.
The main aim of this work is to introduce a statistical approach that can
account for all the complex features of the astronomical data and to apply it to 
available metal abundance measurements. 

In Sect.~2 we first convince ourself about the need of revisiting the way
abundance analyses are performed. We therefore adopted a 
Bayesian method (briefly described in Sect.~3, and detailed in Sect.~5) 
that, when applied to simulated data,  is
able to recover the input enrichment history. With the new method, we
analyse real data (presented in Sect.~4) of 130 clusters observed with
Chandra or XMM. Results based on these (real) data are presented in Sect.~6. 
Sect.~7 describes the advantages of the Bayesian approach and
the limitation of their current implementation. Sect.~8 summarises
the results. Finally, the Appendix provide
the code needed for the revised fitting method.

We adopt $\Omega_m=0.3$, $\Omega_\Lambda=0.7$, and
$H_0=70$ km s$^{-1}$ Mpc$^{-1}$ when computing
ages from redshifts.

\section{The standard analysis}

Iron abundance determination is usually performed 
using XSPEC (or other X--ray spectral fitting programs). XSPEC 
(Arnaud 1996) is a code that returns the likelihood
$\mathcal{L}$ of the parameters describing the
object spectrum, given source and background data, by fitting model
spectra of different metallicity and temperature to the data. 
Users often report a likelihood summary, for example, 
$Z \,\,^{\Delta^+}_{\Delta^-}=0.3^{+0.5}_{-0.3}$, where $Z$ is
the maximum likelihood (mode) value, 
and $Z \,+\Delta^+$ and $Z \,-\Delta^-$ are the points where
the likelihood is lower than its maximum by a given factor
(e.g. $\Delta \chi^2 = 2 \Delta \log \mathcal{L} = 1$).
XSPEC allows fitting, or freezing, other parameters, such as
the cluster redshift, temperature or the Galactic absorption.

When the evolution of the iron abundance is of interest, one should
combine measurements of different clusters. Several
works (e.g. Balestra et al. 2007, Anderson et al. 2009)
combine the information contained in the whole data set in a two
step procedure. First, clusters are grouped by redshift. In
each redshift bin, all cluster data are 
fitted simultaneously, with Galactic absorption and redshift 
fixed at their measured values, both
temperatures and metal abundances 
are free parameters, but temperatures are fitted independently 
for each cluster, while 
the metal abundances are tied (i.e. forced to be the same) across clusters. 
In the
second step, the derived iron abundances are fitted vs redshift, and 
the parameters describing their evolution are found. 

To check whether this standard methodology 
is able to recover the true metal abundance history,
we generate simulated data with known input
parameters. More precisely, we generate simulated Chandra spectra 
of the 114 clusters in Maughan et al. (2012), i.e. having the same
redshift, temperature, and luminosity as the original clusters and
the same exposure time as the original data.  Since clusters
are well known to have a spread of metal abundance (e.g. De Grandi \& Molendi
2001),
the individual metal abundance of each cluster is taken to have a
scatter around the mean relation
$Z_{Fe}/Z_{Fe,\odot}=0.46-0.33 z$, taken from  
Anderson et al. (2009). The adopted amplitude of the scatter, 18\%,
is derived 
from the analysis of 130 real cluster data described in following sections. 
The adopted background and Chandra response matrix
are taken from observations (ObsID 10461 and 12003) used
in Andreon, Trinchieri \& Pizzolato 
(2011)\footnote{Of course, to check the methodology, it would be acceptable
to also use the response matrix and background of any other 
telescope (e.g. XMM).}. 
The generated simulated cluster spectra are then fitted 
following the standard practice (e.g. Balestra
et al. 2007; Maughan et al. 2008) described above. 
We took the same redshift bins as
adopted in Maughan et al. (2008) and Anderson et al. (2009).
The derived metal abundance are fitted vs redshift
using a $\chi^2$ procedure
(e.g. Balestra et al. 2007; Anderson et al. 2009) to determine the
parameters describing the trend between metal abundance and redshift.

This standard methodology returns as the best--fit relation:
$Z_{Fe}/Z_{Fe,\odot}=(0.497\pm0.015)-(0.463\pm0.015) z$. It is fairly different 
in slope from the input relation, $Z_{Fe}/Z_{Fe,\odot}=0.46-0.33 z$
because the two slopes differs by $>8 \sigma$. Furthermore, 
the input parameter values are at $\chi^2 = \chi^2_{min}+13$, where 
$\chi^2_{min}$ is the minimum $\chi^2$. Other difficulties with the standard
analysis are explained in the following sections.

\section{The Bayesian approach in a nutshell}

Very often we know something about the parameter $\theta$ that we want to
study.  For example, we may know its order of magnitude.  This knowledge may be
crystallised in a mathematical expression, $p(\theta)$, stating that, for
example, some values are more probable than others according to a specific
expression $p(\theta)$ (e.g.  $p(\theta)\propto e^{-\theta}$). This probability
distribution is called the prior. Observations return a value, $obs\theta$ and
an  error, or, to be more precise, how likely observing $obs\theta$ is when the
true value is $\theta$, $p(obs\theta|\theta)$.  This probability distribution is
named the likelihood, i.e. a measure of how variable the output $obs\theta$  is
when the input value is  $\theta$. The likelihood (``instrument calibration") is
certainly interesting, but we may perhaps want to know which values of $\theta$
are more probable after having observed $obs\theta$, i.e. to derive the
probability of $\theta$ given $obs\theta$,  $p(\theta|obs\theta)$. The Bayes
theorem tell us that to obtain the latter, called the posterior probability
distribution, we need to
multiply the prior and the likelihood (times a constant of no interest in
parameter estimation).  For a deeper understanding of Bayesian methods, one may
consult, for example, Gelman et al. (2003), and for astronomical
introductions,  Trotta (2007), Andreon \& Hurn (2010, 2011), and Andreon (2010b,
2012), among others.

In this work, we adopt this Bayesian approach. When applied
to the simulated data above, it returns
$Z_{Fe}/Z_{Fe,\odot}=0.45\pm0.02-(0.31\pm0.04) z$, well within $1 \sigma$ from
the input trend. As detailed below, it improves upon 
previous analysis because it operates correctly on likelihoods.
We therefore adopt it for analysing the real data of 
130 clusters, as described in the following sections in detail. 

Readers not interested
in the fitting details may skip Sects. 4 and 5, where we account for 
non--independent and non--Gaussian data, for the existence of an intrinsic
diversity in metal abundances, for mass--dependent selection
effects, and for systematic differences between Chandra-
and XMM- derived metal abundances.

\begin{figure}
\centerline{%
\psfig{figure=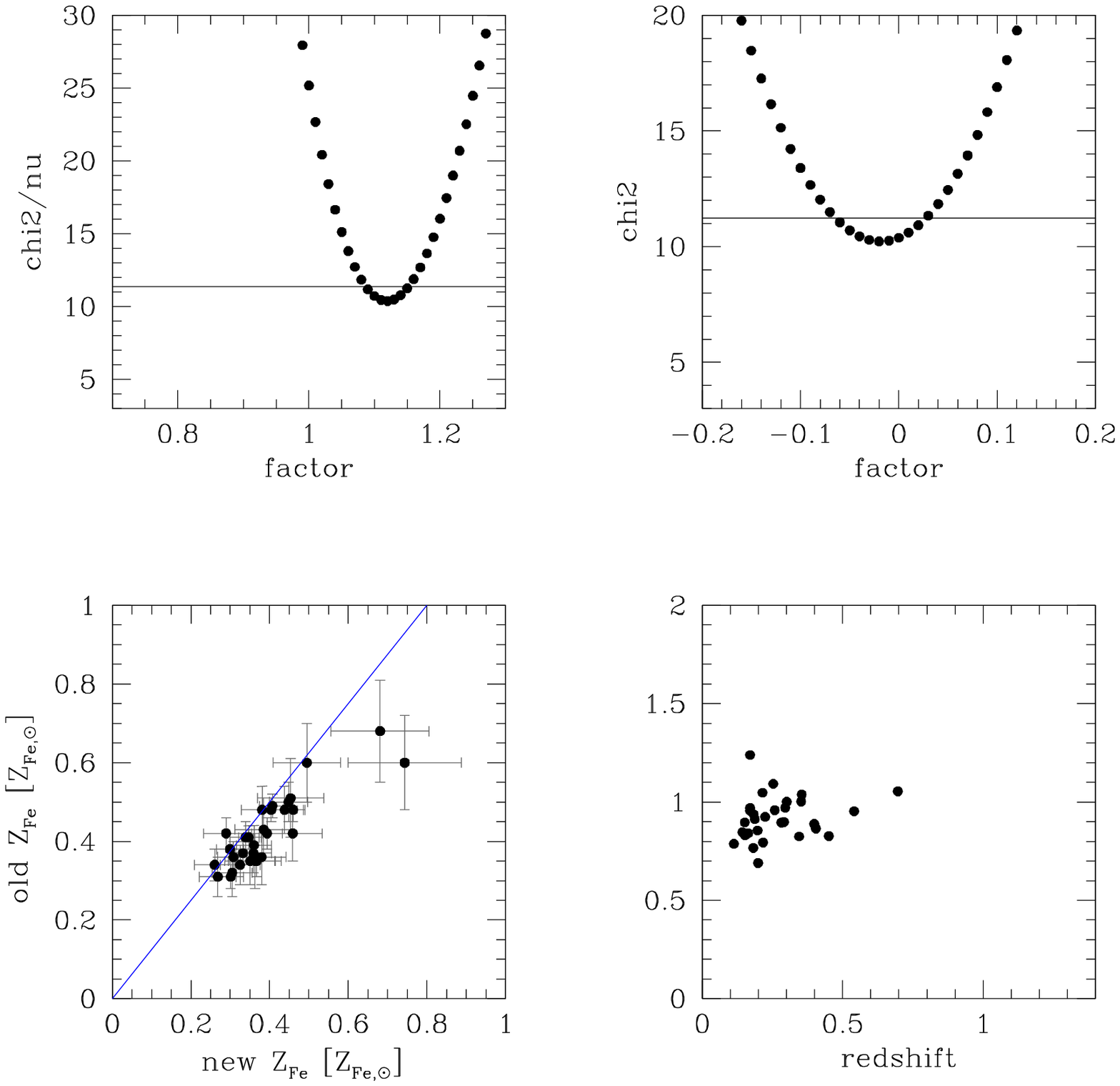,height=6truecm,clip=}%
}
\caption[h]{Old and new Chandra values of metal abundance for the quartile
with the smallest percentage errors on metal abundance. The line indicates 
the one--to--one relation.
}
\end{figure}

\section{The cluster sample \& the new metal abundance measurements}

Our sample is composed of two subsamples: 
114 galaxy clusters at $0.1<z<1.2$ observed with Chandra ACIS-I (Maughan et al. 2008) and
29 galaxy clusters at $0.3<z<1.3$ observed with XMM (Anderson
et al. 2009). For consistency with the Anderson et al (2009) 
analysis, the central region of the clusters is not
excised in Chandra measurements. 

The samples we studied, as well as the two starting lists, 
are heterogeneous collections without any known selection
function being basically, what is available in the Chandra and XMM archives.
The lack of a known selection function is the major limitation of current
samples with available metal abundance measurements, hence also 
of our work, although we control for mass--dependent selection effects, as
described in Sect.~5.4.

Maughan  et al. (2008) present metal abundance and temperature values for his
sample using the CALDB version 3.2.3. The data have
been reprocessed (Maughan et al. 2012) with the up--to--date Chandra calibration,
mainly improving the Chandra mirror effective area
and the ACIS contamination model. In the current work
we also revisit the XSPEC settings
appropriate for metal abundance determination (Sect.~5.7).
These newly derived metal abundances are used here and are listed in Table 1.

Figure 1 compares old and new Chandra metal abundances for the quartile
with the smallest percentage error on metal abundance (to
limit crowding). From the analysis of the whole sample, we
found that metal abundances derived 
with up--to--date calibrations tend to be $12 \pm 3$ \% larger than old values.
A $\chi^2$ analysis returns 
no evidence of a redshift--dependent correction, with
a 68 \% confidence interval of $[-0.06,0.03]$ 
on the the redshift--dependent term.

Fe abundances are normalised to solar abundances in Anders \& Grevesse (1989).
In particular, the solar abundance of iron atoms relative
to hydrogen is $4.68 \ 10^{-5}$.

\section{The fitted model}

As emphasised by the simulation in Sect.~2, the key ingredient for a trustful
determination of the Fe abundance history with current data lays in 
the statistical aspect of the analysis, and this guides our 
presentation.

\subsection{Duplicate clusters}

Thirteen clusters appear in both the Maughan et al. 
(2012) and Anderson et al. (2009) lists,
which make the data dependent. The lack of independence is
quite dangerous if not accounted for, for example, if one computes
the width (intrinsic scatter, dispersion) or the mean
of a distribution (of abundances,
for example) in which some elements are listed twice.

We want independent data and, at the same time, want to make full use of 
the available information. Therefore,
we remove duplicates from the list of fitted data (specifically,
we keep the data set that provides smaller errors on metal abundances), 
and use the removed data in Sect.~5.5 to derive the prior on metal 
abundance systematic. 

The problem of duplicate clusters has already occurred in previous analysis
(e.g. Anderson et al. 2009, Balestra et al. 2007), but is
first addressed here as far as we are aware.

\subsection{Enrichment history}

Metal abundance, $Z_{Fe}$, must be positive at all redshifts.
Some authors (e.g. Anderson et al. 2009) choose to fit
the evolution of the metal abundance with a linear relation, which
may lead to unphysical results. For example,
the best--fit in Anderson et al. (2009) 
crosses $Z_{Fe}=0$ at $z\sim1.2$ i.e. within the
range of the data analysed by them. This is clearly unphysical,
also considering that clusters
at higher redshift exist
(e.g. Stanford et al. 2005, 2009; Andreon et al. 2009, etc.), 
and  for these the best--fit relation predicts negative 
metal abundances. 
We chose to fit metal abundance measurements by a function
that can never cross the physical boundary $Z_{Fe}=0$. It is
also plausible that the Fe abundance was zero at
the time of the Big Bang. 
Therefore, we adopted an exponential function
for the enrichment history parametrised as
\begin{equation}
f(t_i) = \frac{Z_{Fe,z=0.2}}{1-e^{-11/\tau}}(1-e^{-t_i/\tau})
\end{equation}
where $\tau$ is the characteristic
time (in Gyr) and $t_i$ is
the Universe age (in Gyr) at the redshift of the $i^{th}$ cluster.  The denominator
in Eq.~1 was chosen to have $Z_{Fe,z=0.2}$ as second parameter, i.e. the Fe 
abundance (in solar units)
at $z=0.2$, a redshift sampled well by the data used.
This choice simplifies the interpretation of the results.
The $\tau$ parameter regulates when enrichment
occurs: early in the Universe history (small $\tau$'s) or gradually over time
(large $\tau$'s). Two extreme enrichment histories are depicted in
the right hand panel in Fig.~4.  Our choice of using an exponential
function to describe the evolution of the metal abundance is mathematically 
equivalent to assuming a linear function on the log of the metal abundance: 
$Z\propto e^{-t/\tau} \equiv \log Z \propto -t/\tau$. 

With this model, metal abundance is always
positive. Our modelling of the Fe enrichment assumes that
the enrichment starts at the Big Bang. While one may argue that 
enrichment starts at a later age, this is irrelevant for our modelling
as long as the data do not sample very early enrichment phases
(i.e. $t \ll 4$ Gyr, or $z\gg 1.5$). In the future, when observations will
reach the epoch of first enrichment, we will need to replace Eq.~1 with a
more complex formula, for example one with two characteristic times to account
for an initial
enrichment by core collapse supernovae followed by a metal production
spread on longer time scales.  This operation is very easy to implement,
as detailed in the Appendix.

\subsection{Intrinsic scatter}

Galaxy clusters show a spread of Fe abundance 
at the very least because cool--core and not cool--core clusters have
different metallicities (De Grandi \& Molendi 2001). The scatter
may also be due to chemical inhomogeneities and abundance gradients 
within clusters,  differences in cluster star formation histories, 
differences in cluster metallicities (which affect the chemical 
enrichment in as much as the chemical ``yields" depend on metallicity 
e.g. Woosley \& Weaver 1995), different ages of the stellar populations 
in clusters, and different cluster masses. Independently of the
physical source of the spread, the presence of an
intrinsic scatter implies that the information content of a single measurement is
lower than indicated by the error, especially when the latter is comparable to, or is smaller
than, the intrinsic scatter. The intrinsic scatter acts as a floor: the
information content of a measurement is not better than the intrinsic scatter,
no matter how precise the measurement is. Therefore, above some signal--to--noise ratio, 
the information
content of a single measurement no longer increase; in other words, two
measurements with error $\sigma$ smaller than, or comparable to, the
intrinsic scatter are better than one with error
$\sigma/\sqrt{2}$.
Therefore, in the presence of intrinsic scatter,
measurements cannot be combined 
using errors as weight: the error derived from the simultaneous fit
(as in the standard analysis) will be underestimated because it ignores the
metal abundance spread. Furthermore, 
the task of simultaneously fitting spectra
becomes prohibitive if the intrinsic scatter is not known a priori 
because the weight to be used is unknown, 
exactly as in Fe abundance measurements. 
Furthermore, unless individual spectra all have 
the same S/N, the best--fit 
metal abundance will be biased toward the metal abundance of the spectrum 
with the highest S/N. Finally, if 
an intrinsic scatter is not allowed, a
redshift trend may be overly driven by 
a single high S/N measurement, as noted by Baldi et al. (2011), who
note the dramatic effect of including, or removing, 
a very low upper limit in a fit where the intrinsic scatter
is not allowed. 

Instead of fitting spectra of clusters  
with a single value of Fe abundance forced to be the same across
clusters, as
in the standard analysis, we do a simultaneous analysis 
of all the individual spectra allowing Fe abundances to differ
from cluster to cluster and inferring the intrinsic scatter at the same time.
Because metal abundance may evolve and has a 
possible temperature dependence, the intrinsic scatter should be fitted
at the same time as other parameters.
We model the distribution of Fe abundances
as a log--normal process of unknown intrinsic scatter, 
\begin{equation}
Z_{Fe,i} \sim log \mathcal{N} (ln f(t_i) , \sigma^2_{intr}) \quad .
\end{equation}

Expressed in words, the Fe abundance of the $i^{th}$ clusters, $Z_{Fe,i}$ 
shows a log--normal intrinsic
scatter  $\sigma_{intr}$, around the median value, $f(t_i)$. 
Of course, a Gaussian scatter in $Z_{Fe,i}$ is precluded by
the positive nature of the Fe abundance.
The tilde symbol reads ``is distributed as" throughout.

Our adoption of a log--normal scatter removes the major limitation of
previous analysis, namely the tension between data (that require
a spread) and the adopted fitting model (that assumes a unique metal abundance
value at a given redshift). 
The statistical name for this tension is misspecification, and we  
quantify its amplitude in Sect.~6.2.
The choice of a log--normal scatter in $Z_{Fe}$, i.e. a Gaussian scatter
in $\log Z_{Fe}$, is motivated by its being the simplest solution to break
the previously adopted assumption of no scatter. With data of adequate quality,
the shape of the distribution itself may be inferred from the data; however,
this is precluded by the current samples. In Sect.~7.3 we test our
log--normal assumption by adopting instead a Student's $t$ distribution.

\subsection{Controlling for temperature}

The Fe abundance might depend on cluster mass (e.g. Balestra et al. 2007;
Baumgartner et al. 2005).
If neglected, this dependence induces
a bias in determining the evolution of the Fe abundance 
unless the studied sample is a
random, redshift--independent sampling of the cluster mass function. 
For example, if the average mass 
of clusters in the sample increases with redshift and the Fe abundance 
increases with temperature $T$, one may observe
a spurious Fe abundance tilt (increase) with redshift. Other combinations
of dependences are potential sources of a bias, such as 
a decreasing metal abundance with increasing $T$. Among these combinations, we 
should also consider
those that include variations in the mass range sampled at a given redshift
(e.g. lower redshifts sampling a wider cluster mass range). 
To summarise, given
the uncontrolled nature of the available samples, one must at the very
least control
for $T$ (i.e. a mass proxy) in order to avoid the risk of mistaking
a mass dependency for an Fe abundance evolution.
Even if data are unable to unambigously determine a $T$ trend, 
controlling for $T$ allows a trend to be there as much as
allowed by the data and not to overstate the statistical significance
of a redshift trend.

We control for mass by allowing
the Fe abundance, $Z_{Fe}$, to depend on $T$.  We adopt a power law 
relation, following Balestra et al. (2007), 
between metal abundance and temperature, $Z_{Fe} \propto T^\alpha$.
Since clusters are at different redshifts, and $Z_{Fe}$ is possibly evolving,
we need to fit both the $T$ dependency and evolution at the same time.

\begin{figure}
\centerline{%
\psfig{figure=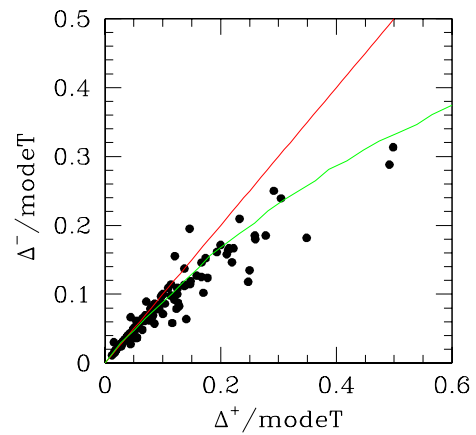,height=6truecm,clip=}%
}
\caption[h]{Determination of the shape of the likelihood function of
temperature. Measured values (points) and expectations for a normal 
(red solid line) and a log--normal (green curve) 
likelihood.   
}
\end{figure}

\subsection{Metal abundances systematics}

Metal abundances may show some systematic differences when
derived by two different teams using data taken with
two different X--ray telescopes (Chandra and XMM) analysed with
similar, but not identical, procedures. In particular, based on
13 common clusters, XMM abundances (derived by Anderson et al. 2009)
are $0.77\pm0.065$ times those measured (by us) with Chandra.

We account for this systematics by allowing metal abundances measured by
different telescopes to differ by a multiplicative
factor as large as allowed by the data. Observationally, the factor 
is constrained by measurements of both telescopes having to
agree after the multiplicative scaling. Of course, to not 
mistake systematics with differences due to the intrinsic scatter or
evolution, dependence on all three parameters have to be accounted for. Therefore, 
systematics have to be inferred at the same time as the other parameters. 
Expressed mathematically, we only need to introduce a quantity, $tid$, that
takes the value of zero for the Chandra data, and one for the XMM data (a convention, 
but one may choose to do the reverse), and
multiply metal abundances by the factor, $1+fact*tid$, to bring all measurements on a common
scale (Chandra, with our convention):
\begin{equation}
Z_{Fe,i,cor} = Z_{Fe,i} * (1+fact * tid_i) * T^{\alpha} \quad .
\end{equation}
The $T^{\alpha}$ term is there to control for $T$, i.e. to account 
for a possible dependence of the Fe abundance on $T$ (Sect.~5.4).

\subsection{Temperature likelihood}

X--ray temperature errors are often asymmetric; i.e., temperatures are 
usually quoted in
the form $modeT^{+\Delta^+}_{-\Delta^-}$, with $\Delta^+
\neq \Delta^-$.  As mentioned in the introduction,
$\Delta^\pm$ are the points where
the likelihood is lower than its maximum by a given factor (see Avni 1976, Press
et al. 1986, or the Sherpa or XSPEC manual for the numerical values to
be used).
A Gaussian function is, of course, symmetric. Therefore, the temperature
likelihood cannot be a Gaussian. We adopt the symbol $modeT$ to emphasise that
the quoted value is the maximum likelihood value (i.e. the mode). A plot of the
temperature likelihood of clusters (e.g. Fig.~12 in Andreon et al. 2009) reveals
the likelihood asymmetry and suggests that it can be described with a 
log--normal shape. The log--normal
distribution has the welcome property of being bounded in the positive
part of the real axis, i.e. avoids
negative $T$. 

Our suggestion of a log--normal temperature likelihood can be
tested with our large sample. Figure 2 shows
$\Delta^+/modeT$ vs $\Delta^-/modeT$. For a log--normal function, the 
points should follow the curved (green) line. If the likelihood were
Gaussian, the locus (straight line) $\Delta^+/modeT=\Delta^-/modeT$ should be
followed.  Figure 2 shows that the log--normal likelihood captures the data
behaviour. Simple algebra, together with the knowledge of the definition of
$\Delta^+$ and $\Delta^-$ (see XSPEC or Sherpa manuals), allows us  to
analytically compute the likelihood parameters. For a log--normal model with
location $\mu$  and scale $\sigma$, we find
\begin{eqnarray}
\sigma_\pm=\pm \ln (modeT\pm\Delta^\pm) \mp \ln (modeT) \nonumber \quad .
\end{eqnarray}
Using real data, $\sigma_+$  and $\sigma_-$ slightly differ
because every measurement, including
$\Delta^\pm$ values, is subject to errors and rounding. We thus take
the average of $\sigma_\pm$ as the value of $\sigma_{T,i}$. 

To sum up, the whole section may be summarised by the mathematical
expression:
\begin{equation}
modeT_i \sim log\mathcal{N}(\ln T_i,\sigma^2_{T,i}) \quad ,
\end{equation}
which allows us to account for asymmetric temperature errors.
While our statement of the non--normal $T$ likelihood is certainly not new, 
to our knowledge this is the first time the non--normal $T$ likelihood has been 
implemented in a regression fitting involving $T$.

\begin{figure}
\centerline{%
\psfig{figure=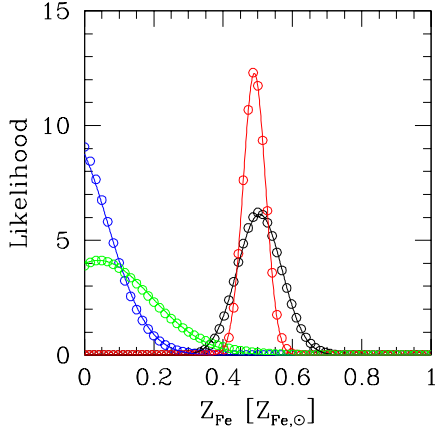,height=6truecm,clip=}%
}
\caption[h]{Abundance likelihood function of four clusters, chosen to illustrate
the likelihood shape in various cases. The open circles show the likelihoods
derived with XSPEC, whereas the solid curves present a Gaussian approximation of them. 
}
\end{figure}

\subsection{Fe abundance likelihood}

Inspection of plots of the metal abundance likelihood (four examples are
shown in Fig.~3, computed using XSPEC) shows, in agreement with Leccardi 
\& Molendi (2008), that the 
metal abundance likelihood has a Gaussian shape. In formulae
\begin{equation}
modeZ_{Fe,i} \sim \mathcal{N}(Z_{Fe,i,cor},\sigma^2_{Z_{Fe,i}}) \quad .
\end{equation}
This implies that a negative $modeZ_{Fe,i}$ can be found, most often
for low S/N measurements.

The parameters of the equation above are determined by fitting 
the data with XSPEC. We emphasise, however, that the input numbers of
this equation, $\sigma_{Z_{Fe,i}}$ and $modeZ_{Fe,i}$, are not, generally speaking,
the standard XSPEC output numbers because XSPEC uses different definitions 
of the likelihood location and width. 
 
The mode of the likelihood on the unrestricted range $Z_{Fe} \in [-\infty,\infty]$
is $modeZ_{Fe,i}$.
We accept negative
$modeZ_{Fe,i}$ values as a way to describe the likelihood wing at $Z_{Fe,i}>0$ when
the latter shows no maximum there (e.g. the leftmost likelihood in Fig.~3). 
Negative
$modeZ_{Fe,i}$ indicate that the measurement has such a low S/N that the
likelihood has no peak in the physical range ($Z_{Fe,i}>0$), and 
the data only offer an upper limit to the cluster metal abundance. 
We emphasise that in the Bayesian approach inferences come from the
posterior, not from the likelihood alone. Since the abundance posterior distribution 
is only non--zero for $Z_{Fe}>0$ (because of the prior,
Eq.~6) every point estimate of the cluster abundance (e.g. posterior mean, median)
is always positively defined, as good sense requires, even
when $modeZ_{Fe,i}<0$.
XSPEC's 
$maxZ_{Fe,i}$ is the mode of the likelihood in a restricted range 
(set at $Z_{Fe}\ge 0$ by default). It differs from $modeZ_{Fe,i}$
when the likelihood has no peak at $Z_{Fe,i}>0$ (e.g. the left--most likelihood
in Fig.~3).

The usual width of the Gaussian is $\sigma_{Z_{Fe,i}}$. For the author's opinion, it is the most 
straightforward estimate of the metallicity uncertainty.
XSPEC instead quantifies the likelihood width by
$\Delta^\pm$ values, determined by the points where
the likelihood is lower than its maximum by a given factor. If the peak is far
away from the zero boundary (mathematically: if $modeZ_{Fe,i} -\sigma_{Z_{Fe,i}} > 0$), 
as in the two
rightmost likelihoods in Fig.~3, the two estimates of the likelihood
width coincide. However,
these estimates may differ widely, as in the left--most
likelihood in Fig.~3, where
$modeZ_{Fe,i}+\sigma_{Z_{Fe,i}}\approx 0$. In such cases, 
$\Delta^+$ value is low, but 
the error $\sigma_{Z_{Fe,i}}$ is large
(the data only offer an upper limit to the cluster metal abundance.)
Therefore, one should not overlook the
difference between $\Delta^\pm$ and $\sigma_{Z_{Fe,i}}$.
If $modeZ_{Fe,i} -\sigma_{Z_{Fe,i}} < 0$, 
then $\Delta^-$ differs from $\Delta^+$, 
even for a Gaussian likelihood (e.g. for the two leftmost
likelihoods displayed in Fig.~3). This situation is quite usual for
cluster metal abundances, and in fact asymmetric $\Delta^\pm$ values
are often quoted (e.g. $0.3^{+0.5}_{-0.3}$). 

The two definition sets may be made identical by removing
the XSPEC default positivity constraint to abundances. Alternatively,
one may note that $\sigma_{Z_{Fe,i}}=\Delta^+$, except when $maxAb=0$.
We adopted this property for Anderson et al.
(2009) abundance measurements.
When XSPEC (with the positivity constrain set on) instead returns $maxAb=0$ (e.g. CLJ0522-3625), 
the XSPEC positivity constraint has to be removed, and we accept negative
$modeZ_{Fe,i}$ values. This choice allows us 
to analyse samples, like the studied one that include both upper limits and precisely
determined values. In this way,
including upper limits is automatically accounted for in our determination of the metal 
abundance history.

We verified that if one 
overlooks the difference between $\Delta^\pm$ and $\sigma_{Z_{Fe,i}}$, 
the trend between metal abundance and
redshift is tilted because these noisy values occur only at high redshift.

\medskip

Table 1 lists the cluster ID, 
the redshift $z$, the temperature $modeT_i$, and its 
$\sigma_{T,i}$, the metal abundance $modeA_i$ and its $\sigma_{Z_{Fe,i}}$, and 
indicates whether
the data comes from Chandra or XMM ($tid$). The table has 130 lines in
its electronic version. About 70\% of the listed temperature or
metal abundance values and errors have been newly derived
(the remaining values are taken from Maughan et al 2012 or Anderson et al.
2009). We emphasise, as explained above, that negative values
for $modeZ_{Fe}$ in Table 1 are correct (whereas every posterior
estimate of $Z_{Fe}$ is positive)
and are meant to describe the Gaussian likelihood wing at $Z_{Fe,i}>0$.
Forcing them to be positive would spuriously 
bend the trend between metal abundance and redshift.

\begin{figure*}
\centerline{%
\psfig{figure=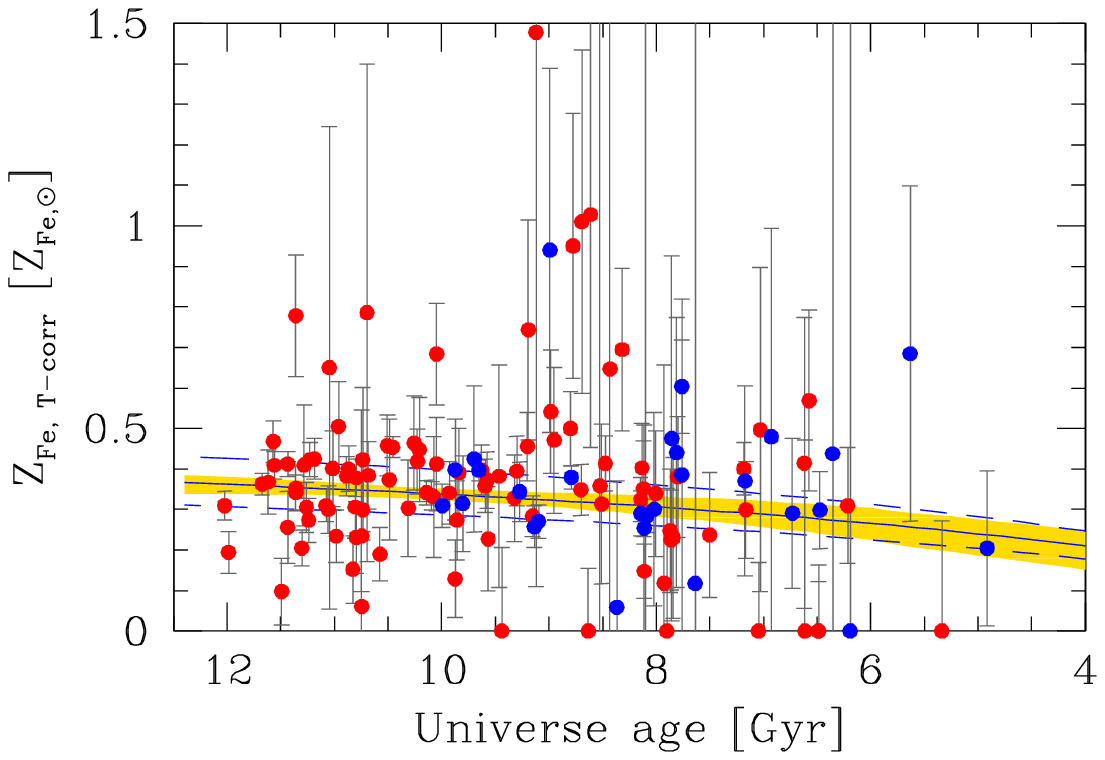,width=9truecm,clip=}%
\psfig{figure=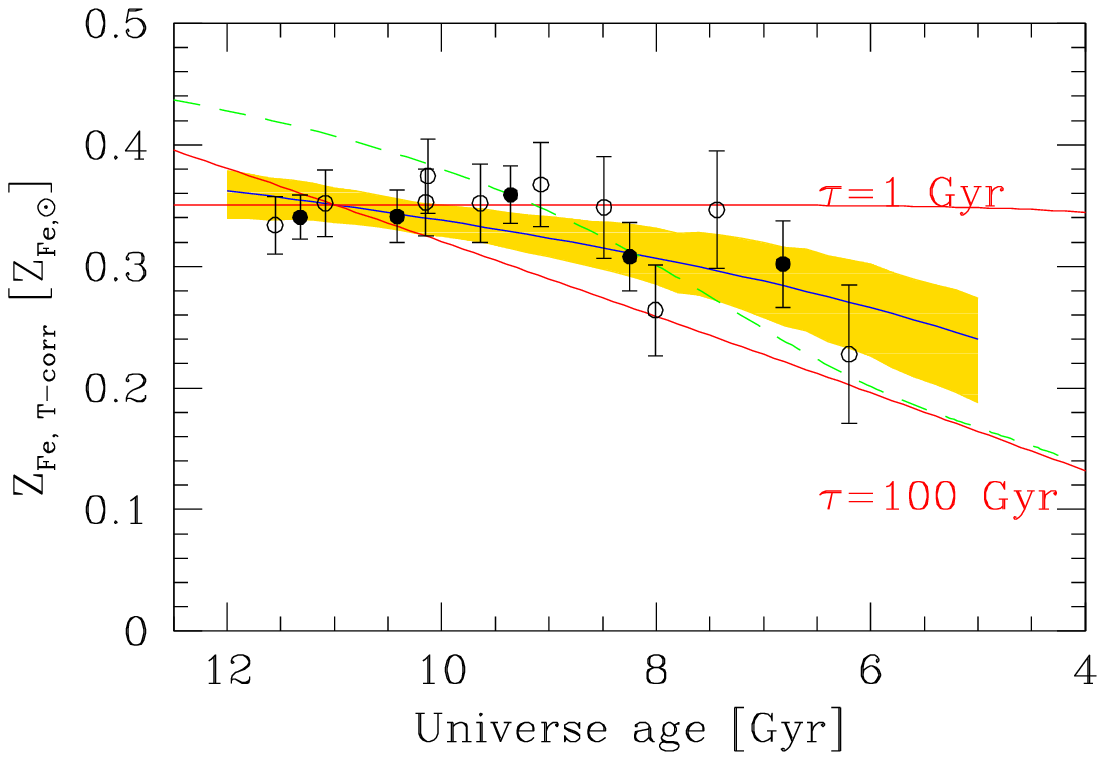,width=9truecm,clip=}%
}
\caption[h]{Metal abundance, on the Chandra scale, vs Universe age. {\it Left Panel:} 
Observed values of metal abundance and errors are
corrected for the $T$ dependence and for the Chandra vs XMM systematic as determined
by a simultaneous fit of all parameters. Red (blue) circles refer to Chandra (XMM)
measurements. 
Points indicate the maximum a posteriori, whereas error bars represent
the shortest 68\% probability intervals.
The solid line marks the mean fitted relation between  metal abundance and
redshift, while the dashed line
shows this mean plus or minus the intrinsic scatter $\sigma_{scat}$. The shaded 
region marks the 68\%
highest posterior credible interval for the regression.
The distances between the data and the mean model are due in part to the
measurement error and in part to the intrinsic scatter. {\it Right Panel:} Solid
points are metal abundances, corrected as in the left panel, but binning in 
5 (solid points) or 10 (open points) redshift bins. The solid line
and shading are as in the left panel. Two extreme enrichment histories
(red lines) and the Ettori (2005) model (green dashed line) are also plotted .}
\label{fig:fig2}
\end{figure*}

\subsection{Priors}

At this point, we have the data and we described the mathematical link between
the quantities that matter for our problem (Eqs. 1 to 5). To complete the
analysis, we now specify what else we know about the parameters. 
Except for metal abundance systematics, we assume we know very little about the parameters; 
i.e.,
we adopt for all quantities priors  wide enough to certainly include
the true value, but not so wide to include unphysical values.
For the systematics on metal abundance
we adopt as prior the result of our metal abundance comparison (Sect.~5.5)
for the 13 clusters in common between Chandra and XMM lists:  $0.77\pm0.065$.

\begin{table}
\caption{Id, Redshift, metal abundances, temperatures, and their $\sigma$ values, and data source
(zero for Chandra, one for XMM--Newton).}
{
\scriptsize
\begin{tabular}{l r r r r r r}
\hline
id & $z_i$ & $modeZ_{Fe,i}$ & $\sigma_{Z_{Fe,i}}$ & $modeT_i$ & $\sigma_{T, i}$  &  $tid_i$\\
  & & [$Z_{Fe,\odot}$] & [$Z_{Fe,\odot}$] & [keV] &  & \\
\hline
MS1008.1-1224 & 0.301 & 0.41 & 0.11 & 5.00 & 0.06 & 0 \\ 
CLJ1113.1-2615 & 0.725 & 0.51 & 0.41 & 3.80 & 0.18 & 0 \\ 
RXJ1716.9+6708 & 0.813 & 0.55 & 0.22 & 6.30 & 0.14 & 0 \\ 
A2111 & 0.229 & 0.23 & 0.13 & 6.40 & 0.09 & 0 \\ 
A697 & 0.282 & 0.38 & 0.07 & 9.80 & 0.05 & 0 \\ 
\hline
 \hline 	
\end{tabular}
} \hfill\break 
This table has 130 lines in its electronic version.
\end{table}    

Specifically,
the prior of the mean Fe abundance at $z=0.2$, $Z_{Fe,z=0.2}$
is taken as
uniform between 0 and 1 $Z_{Fe,\odot}$; i.e., the  mean Fe abundance may 
take any value in this  wide range  
with no preferred value. Similarly, for the true value of the cluster temperature, 
we took an uniform distribution over a wide range 1 to 20 keV, 
which generously 
includes all plausible values for the true temperature of the studied
clusters. Their observed values range from $2.4$ to $14.7$ keV. Similarly, 
the prior of the  intrinsic scatter of Fe abundance, $\sigma_{intr}$ is
also taken as uniform between 0 and 1, a range wide enough to certainly include
the true value. 
The prior on $\tau$ is taken to be uniform between the wide range 1 and 100 Gyr. More extreme 
values give enrichment histories that are
indistinguishable from $\tau=1$ Gyr or $\tau=100$ Gyr  
in the redshift range explored in this work. 
This is the reason we adopted these
values as boundaries of the explored parameter space.
If $\tau \approx 1$ Gyr, the
Fe abundance is constant (see right panel of Fig.~4)
and if $\tau \approx 100$ Gyr, models change almost linearly 
(see right panel of Fig.~4) with age.  
The prior of the power--law dependency of
iron abundance on temperature is a Student's $t$ (uniform on the
angle $a=\arctan \alpha$), as in previous works dealing with 
a slope computation 
(e.g. Andreon et al. 2006, Andreon \& Hurn 2010). 
In formulae the priors are

\begin{eqnarray}
Z_{Fe,z=0.2} &\sim& \mathcal{U}(0,1) \\
\tau &\sim& \mathcal{U}(1,100) \\
\alpha &\sim& t_1  \\ 
fact&\sim& \mathcal{N}(0.77-1,0.065^2) I(-1,)\\
\sigma_{intr} &\sim& \mathcal{U}(0,1) \\
T_i &\sim& \mathcal{U}(1,20) 
\end{eqnarray}
where the $I(-1,)$ operator truncates the likelihood at $-1$, to avoid
$(1+fact)$ taking negative (unphysical) values. 

How much our conclusions depends on the chosen priors is detailed below,
but we can anticipate that the dependence is almost zero.

\begin{figure*}
\centerline{%
\psfig{figure=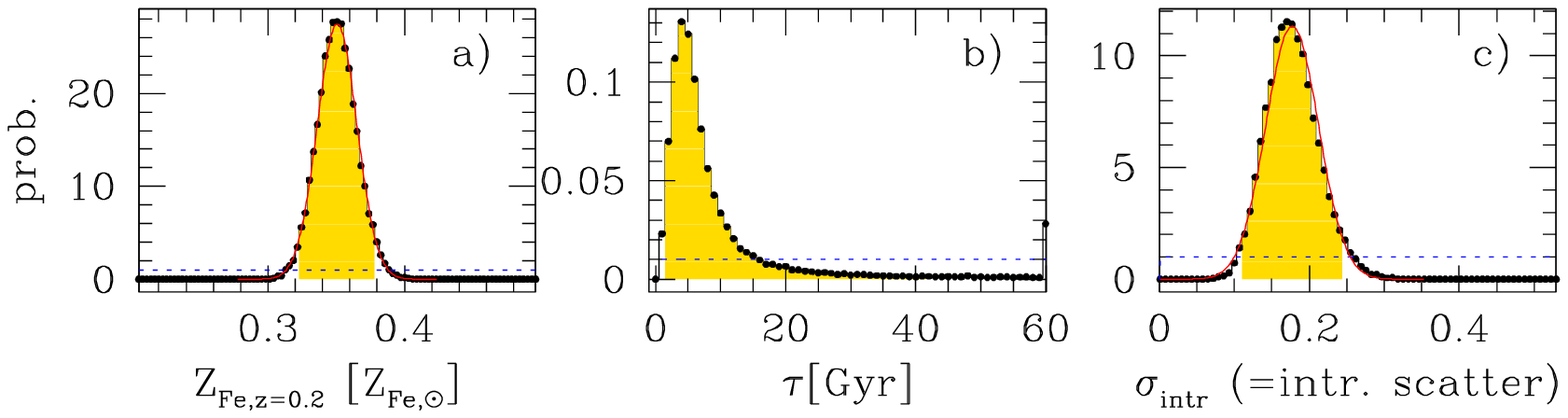,height=3truecm,clip=}\hskip -0.5 truecm %
\psfig{figure=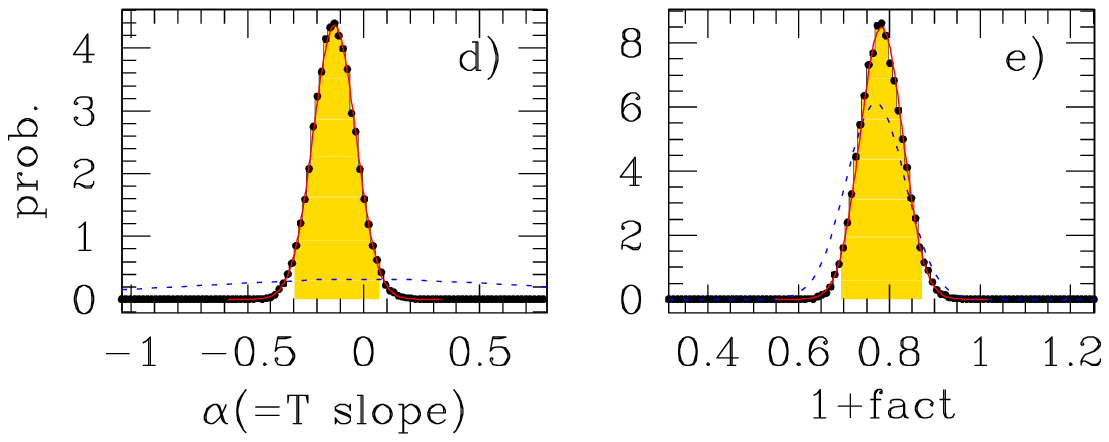,height=3truecm,clip=}%
}
\caption[h]{Probability distribution for the
parameters of the metal abundance vs redshift fit.
The solid circles show the posterior probability distribution as computed
by MCMC, marginalised over the other parameters.
The red curve (when present) shows a Gauss approximation of it.
The dashed curve displays the adopted prior.  
The shaded (yellow) range shows
the 95\% highest posterior credible interval. 
\label{fig:fig3}
}
\end{figure*}

\subsection{Stochastic computation}

At this point, we have the data in the right format (i.e. with $\sigma$'s), 
we have the description of the link between interesting parameters
(Eqs. 1 to 5), and we have specified 
what we know about the parameters before seeing the data (Eqs. 6 to 11).
We now need to use the Bayes theorem and to compute the
posterior probability distribution of the parameters.  
Just Another Gibb Sampler 
(JAGS\footnote{http://calvin.iarc.fr/$\sim$martyn/software/jags/}) 
can return it in form of (Markov Chain) Monte Carlo samplings.
From the
Monte Carlo sampling one may directly derive mean values, standard deviations,
and confidence regions of any parameter or any 
parameter--depending quantity. For example, for a 90 \% interval on $\tau$, 
it is sufficient to
take the interval that contain 90 \% of the $\tau$ samplings.

Readers who are less familiar with Bayesian methods may
just think that we use JAGS to correctly combine likelihoods (data)
in order to extract parameters values.
The JAGS code is given in Appendix.

\section{Results}

\subsection{Parameter estimation}

The result of the fit of
metal abundance and temperature values for the sample of 130 clusters is
summarised in Figs. 4 and 5. 
The left--hand panel of Fig.~4 shows the data, corrected for the $T$ dependence 
and for the metal abundance systematics, 
as determined by a simultaneous fit of all parameters,
the mean fitted relation between metal abundance and
Universe age, 
this mean plus or minus the intrinsic scatter $\sigma_{scat}$, that
turns out to be $0.18\pm0.03$, i.e. 18 \% of the Fe abundance value, and
the 68\%
highest posterior credible interval of the fit. 

An useful approximation of the mean trend is given by
\begin{equation}
\frac{0.35}{1-e^{-11/6}}(1-e^{-t/6}) \quad ,
\end{equation}
with errors given by
$\pm0.02$ at ages greater than 7 Gyr,
and $\pm0.04$ at younger ages.

Figure 5 shows prior and posterior probability distributions of the parameters. The
figure highlights two key points: first, the posterior distribution of all parameters 
is much more concentrated than the prior; i.e., data are highly informative about
these parameters. As a consequence, conclusions  on these parameters do not depend on
the adopted prior (see below for discussion about panel e). Second, the posterior
probability distribution of all parameters except $\tau$ is described well  by a 
Gaussian distribution.

The metal abundance at $z=0.2$, 
$Z_{Fe,z=0.2}$ is fairly well determined: $0.35\pm0.01$ $Z_{Fe,\odot}$
(Fig.~5, panel a). 
The intrinsic scatter in abundance
values, after controlling for $T$, taking 
evolution and systematics into account, is $0.18\pm0.03$ (Fig.~5, panel c).
The posterior distribution of the e--folding time $\tau$
avoids both low (near 1 Gyr)
and high (above 20 Gyr) values and peaks on timescales of 4-6 Gyr. The
(highest posterior) 68 \% probability interval is [1.4-8.6] Gyr.
The flat shape of the right tail of the posterior  occurs because
models with $\tau\approx20$ Gyr show very tiny differences, too small to be
measurable with the current sample. The key point to keep in mind is that
the posterior probability distribution peaks at $\tau \approx 4-6$ Gyr; i.e., 
enrichment histories
completed at high redshift or delayed at late times are not favored 
by the data, a statement that we further quantify in Sect.~6.2. 
While entropy feedback predates cluster formation (Ponman et al. 1999),
metal abundance enrichment is not exhausted at high redshift. 
We emphasise that if metal abundance upper limits were incorrectly dealt with
(and everything else dealt correctly),
shorter e--folding time $\tau$ (more strongly evolving metal abundance histories) 
would be found with increased (spurious) statistical significance.

The slope $\alpha$ of the metal abundance vs temperature scaling 
is certainly not steep, but, apart from that, is poorly 
determined: $-0.12\pm0.09$ (Fig.~5, panel d).
As mentioned, by letting it be as large as permitted by the data
allows us not to mistake a $T$ dependency, joined to a non--random
sample selection, with a metal abundance evolution. It also allows us to
remove a source of concern, because we now 
know that mass--selected effects are minimal, if there are any, for cluster 
samples similar (in cardinality, mass, and redshift range) to our one.
The slope found by Balestra et al. (2007), $-0.47$ is $\sim 3.8\sigma$
away from our determination, but it has been derived using a three
times smaller sample ignoring
the evolution of the iron abundance, the intrinsic scatter, and metal abundance
systematics.

Metal abundances measured with XMM turn out to be $0.78\pm0.045$  times 
those measured with Chandra (Fig.~5, panel e), in agreement 
with (and improving upon) our estimate
of Sect.~5.5 (prior), $0.77\pm0.065$. 
If an uniform prior were adopted for this parameter,
the posterior mean would be $0.80\pm0.07$.
The agreement between the amplitude of the
systematic derived from the 130 clusters alone ($0.80\pm0.07$)
and the one independently derived from the 13 clusters observed
by both Chandra and XMM ($0.77\pm0.065$)
indicates that the measurement of the abundance systematics 
is robustly determined.  The precision
achieved by comparing the metal abundance of 13 clusters with both
XMM and Chandra measurements ($0.065$) is similar to the one inferred from
a ten times larger sample that assumes no metallicity bias between
clusters observed by Chandra and by XMM ($0.07$).

The overall decline in the metal abundance that we observe, also reported in
the right--hand panel of Fig.~4, is slower than in the phenomenological
model of Ettori et al. (2005). The 
offset at low redshift between Ettori et al. (2005) and current metal
abundance measurements is due to the change of the Chandra calibration.
In fact, by repeating
our analysis using the older metal abundances (from Maughan et al. 2008),
we found better agreement 
at low redshift with the Ettori et al. (2005) model. 

\begin{figure}
\centerline{%
\psfig{figure=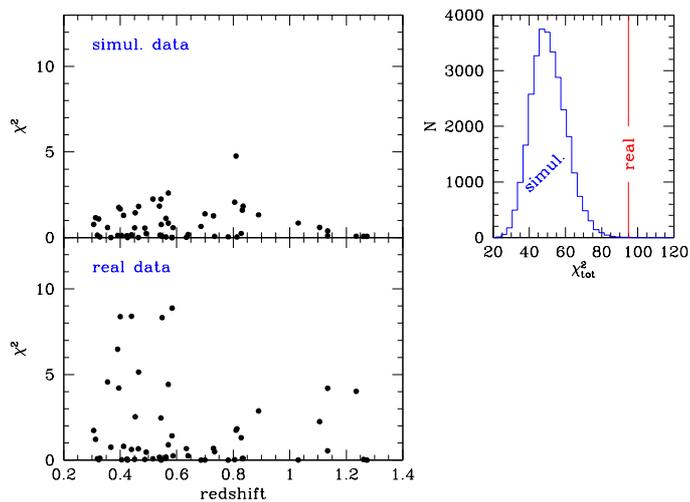,width=9truecm,clip=}
}
\caption[h]{Illustration of how the fitted data reject the
best--fit in previous works. The bottom--left panel shows the 
$\chi^2$ of the individual data points (from Balestra et al. 2007, chosen 
for illustration) vs redshift. The $\chi^2$ is computed using the Balestra
et al. (2007) best--fit. The top left-hand panel shows the same, but for data extracted
from a realisation of the Balestra et al. (2007) fitted model. Real data scatter
much more than simulated data, as quantified in the top right-hand panel:
the total $\chi^2$
of 30000 realisations of the Balestra et al. (2007) model is histogrammed,
together with the total $\chi^2$ of their sample
(vertical line). Either using $\chi^2$ tables (or the usual rule $\chi^2/\nu$,
$\nu\approx54$)
or more sophisticated simulations, the Balestra et al. (2007) data reject 
their best--fit at $>99.9$ confidence.
\label{fig:fig6}
}
\end{figure}

Because of the youthfulness of theoretical (numerical or analytic)
models (see Fabjan et al. 2008 for a careful report about the
limitations of their own
numerical modelling), our observational
constraints on the metal abundance history 
cannot be transformed in physical constraints on cluster parameters, so 
we cannot interpret the observational results as due to 
clumps of enriched gas falling in the cluster centre (e.g. Cora et al.
2008), a recent pollution by
metals previously present in stars (e.g. Calura et al. 2007),  
an enhanced star formation at low redshift, or any other proposed
mechanisms. There are two reasons that prevent us from drawing firm conclusions:
first, different physical mechanisms implemented in models
are able reproduce this specific 
cluster observable, the metal abundance history, and second,
these models are generally not able to reproduce other related
observables (e.g. the $L_X-T$ scaling relation, the stellar baryon fraction, 
the temperature profile, etc.). For example, numerical simulations
produce the right amount of metals, but generate 
ten times too many stars (Andreon 2010).

\subsection{Model missfit and evidence for evolution}

After computing the model parameters that best describe the data, we
also checked whether the adopted model provides a good description of
the data, or  whether it is misspecified, i.e. in tension with the
data.  In the non--Bayesian paradigm, this is often achieved by
computing a p--value, i.e. the probability of obtaining more discrepant
data than those in hand, once parameters are taken at the best--fit
value (i.e. the number returned by  the Spearman, Kolmogorov--Smirnov,
$\chi^2$, F, etc. tests).  The Bayesian version  of the concept (e.g.
Gelman et al. 2004, Andreon 2012) acknowledges that parameters are not
perfectly known,  and therefore one should also explore other values of
the parameters in addition to the best--fit value.   To this end, we
generated simulated data extracted from the model  (i.e. from every set
of parameters, each one occurring with a probability given by the
posterior probability distribution computed in the previous section) 
and counted which fraction of them are more discrepant than the real
data.  As a measure of discrepancy we adopted the modified $\chi^2$, i.e.
one having an estimate of the total (error plus
intrinsic scatter) variance at the denominator. 
This simulation, too, is performed in JAGS
as explained in the Appendix. The simulation accounts for all  modelled
sources of uncertainties (intrinsic scatter, measurement errors, their
non--Gaussian nature, etc.). We performed 30000 simulations, each one
generating 130    measurements of Fe abundance.  We found that 62 \% of
the simulated data sets shows a larger $\chi^2$ than the one of real
data. Therefore, real data are quite common, given the assumed model,
and  our model shows no evidence of misspecification.  This agreement
should not be taken for granted, however. If we adopt the Anderson et
al. (2009) modelling, a linear relation without any intrinsic scatter, 
then we find that their best--fit model is rejected by their data with
$>99.99$ \% probability,  because none of the 30000 simulated datasets
has a larger $\chi^2$.  Similarly, we found that the p--value of the
best--fit model quoted in Balestra et al. (2007), computed using their
same unbinned  data used by these authors, is rejected with $>99.9$ \%
probability. Both p--values have been computed by
performing a simulation, as mentioned above.  
Therefore, the Balestra et al. (2007) and Anderson
et al. (2009) best--fits are poor fits of their data.  Our conclusions
are robust with respect  to the choice of a Bayesian, or non--Bayesian,
p--value. 

\begin{figure}
\centerline{%
\psfig{figure=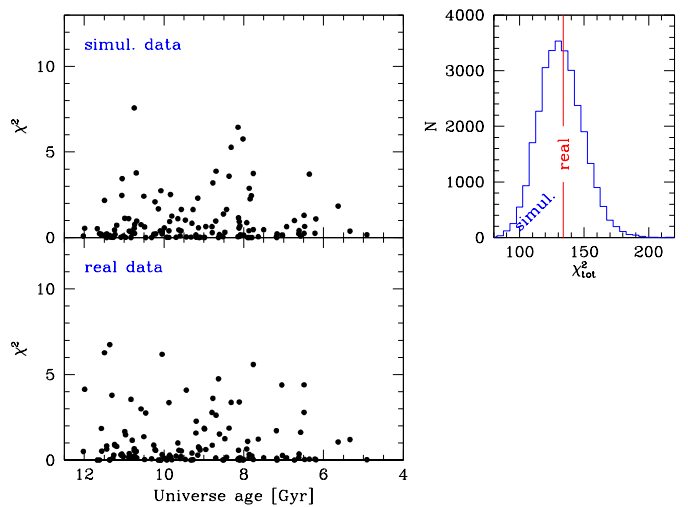,width=9truecm,clip=}
}
\caption[h]{As in Fig.~6, but for current data and fit, using Universe
age instead of redshift as abscissa. Whether using $\chi^2$ 
tables ($\nu\approx130$)
or more sophisticated simulations, the current fit ($\chi^2_{tot}=131$) is 
found to be a good description of the data.
\label{fig:fig7}
}
\end{figure}

In Fig.~6, we show the improvement of the fit 
to the $Z_{Fe}-z$ relation achieved in this paper with 
respect to previous works (Balestra et al. 2007,
chosen for illustration
purposes). 
The plotted $\chi^2$ values are computed using their data and best--fit.
We expect that most of the points have $\chi^2_i\approx 1$, that rarely
$\chi^2_i\ga 9$, and 
that the total $\chi^2_{tot}$ is about the number of degree
of freedom, $\approx 54$,
(the usual $\chi^2_{tot}/\nu \approx 1$ definition of 
a good fit). Instead, the total $\chi^2$ is 94, and 
several points have a $\chi^2$ near $9$. 
Reading from $\chi^2$ tables (e.g.
Bevington 1969), the best--fit can be rejected at
99.9 \% confidence by their data. We do not need to rely on these tables, 
or on the asymptotic behaviour assumed by these tables, but
we can use simulated data sets, whose first realisation 
is shown in the top left--hand panel. We  note
that points with $\chi^2\ga 5$ do not occur and
that, on average, $\chi^2_i$ values tend to be 
smaller than in bottom panel. 
The top right--hand panel quantifies this visual impression by
comparing the total $\chi^2$ of the real sample to the
the distribution of the total $\chi^2$ of 30000 simulated data sets. 
The latter have,
as expected, a typical total $\chi^2$ of $\approx 54$. 
The $\chi^2$ of the observed data is larger than 99.9 \% of the simulated
data. These plots illustrate that the Balestra et al. (2007) data reject
their best--fit at 99.9 \% confidence. This occurs because of 
greater variability of the
Balestra et al. (2007) data than allowed by their fitted model. A similar
plot can be constructed for Anderson et al. (2009).

The full, Bayesian, computation described
above refines this simple analysis by accounting for
other terms, such as the possible temperature dependency, 
allowing for metal abundance systematics, accounting for uncertainties
in fitted parameters, but not changing the conclusions: Balestra et al. (2007) 
and Anderson et al. (2009) best--fits are
poor fits of their data. Figure 7 illustrates, in a simplified way, how
well our model fit the data: deviations from the
mean trend visible in the bottom left--hand panel (i.e. $\chi^2$ of a few) 
are now present in the simulated data set, plotted
in the top left--hand panel and drawn from the fitted model.
The $\chi^2$ of the real data ($131$) is similar to the number of degree of 
freedom ($\approx 130$) 
and quite a  typical value for simulated data, indicating our
fitting model is a good description of the data.

In Sect.~6.1 we mentioned, without any quantification, 
that data do not favour early or late enrichment 
histories. The support of the data to an intermediate  ($3 \la \tau \la 8$ Gyr)
enrichment history $M_0$, over
a late or early ($\tau \ga 8$ Gyr or $\tau \la 3$ Gyr ) 
history $M_1$ 
is given by the ratio of the probabilities
of the two models $p(M_0)/p(M_1)$.
This ratio can be easily computed 
in our case, because it is given by the ratio between
$p(3 \la \tau \la 8)/(1-p(3 \la \tau \la 8))$ and
$\pi(3 \la \tau \la 8)/(1-\pi(3 \la \tau \la 8))$ 
where $p$ is the posterior probability distribution and 
$\pi$ the prior (Congdon 2006, p. 68).  In practice,
we just need to compute the ratio of the integrals below  the solid and dotted
lines in the $\tau$ panel of Fig.~5 at $3 \la \tau \la 8$ Gyr. 
We found that intermediate
enrichment histories are favored with
19 to 1 odds, a moderate evidence on the Jeffreys (1961) scale.
The right-hand panel of Fig.~4, which shows average metal abundances per redshift bins, may
also qualitatively hint at an intermediate metal abundance history.
As discussed below, this evidence should not be read as ``at 95 \%
confidence'' because the evidence scale differs from the statistical
significance usually quoted in many works.

To claim an evolving Fe abundance, other authors (e.g.
Balestra et al. 2007 and Anderson et al. 2009) have
tested whether their unbinned data are described by
a non--evolving Fe abundance by computing a p--value, the tail probability
derived using the Spearman test. Since an
extreme p--value is found, these authors conclude that the model tested,
a non--evolving one, is ruled out. However, this conclusion is hasty,
since the p--value does not identify
what is wrong with the model, but only indicates that
something is wrong.  We can test whether 
the extreme p--value is due to the non--evolving Fe abundance, 
as claimed by these authors, or to some other misspecified
model ingredients. We do this by
computing the the p--value of their best--fit model (which is evolving)
using the very same data as used by the
authors to reject the non-evolving metal abundance. As computed above, 
both the Anderson et al. (2009) and Balestra et al. (2007)
unbinned data reject their best--fit model. 
This confirms the flaw in the reasoning:
the extreme p--value these authors find is mostly driven 
by model misspecifications (e.g. neglecting the intrinsic scatter),
not by the evolution of the metal abundance value.

To summarise, one should not interpret model missfits (a tail probability) 
as evidence of evolution (the odds, a ratio of two probabilities).
We also emphasise that it is easier to get larger proofs for an
evolving metal abundance history by paying less attention to the statistical
analysis. We verified, for example, that an incorrect treatment of metal abundance
upper limits spuriously
reinforces the evidence of the trend (Sect.~5.7). 
Similarly, stacking spectra or fitting trends
ignoring the intrinsic scatter  
leads to underestimated
error bars (Sect.~2), i.e. to overestimate the evidence of evolution.

\subsection{Binning by redshift}

In this section we take a step towards an analysis more typical of astronomical
papers, but, to do this, we need to accept some approximations. 

Some authors may prefer not to impose the functional $t$ dependency in Eq.~1, 
and just inspect the enrichment history found by binning cluster data
in redshift bins. This can be done
easily, it is just a matter of removing Eq.~1 and allowing metal abundances
in different redshift bins to be independent of each other.
The logical link between intervening quantities becomes

\begin{eqnarray}
Z_{Fe,i} &\sim& log \mathcal{N} (ln (meanZ_{Fe}), \sigma^2_{intr}) \\
Z_{Fe,i,cor} &=& Z_{Fe,i} * (1+fact * tid_i) * T^{\alpha} \\
modeZ_{Fe,i} &\sim& \mathcal{N}(Z_{Fe,i,cor},\sigma^2_{Z_{Fe,i}}) \\
modeT_i &\sim& log\mathcal{N}(\ln T_i,\sigma^2_{T,i}) \\
T_i &\sim& \mathcal{U}(1,20) \quad ,
\end{eqnarray}
to which we should add the prior for the newly introduced quantity,
the median metal abundance in the redshift bin, $meanZ_{Fe}$, 
taken to be a priori in a wide range with no preference
of any value over any other:
\begin{equation}
meanZ_{Fe} \sim \mathcal{U}(0,1) \quad .
\end{equation}

We adopt the posterior probability distributions determined in the previous 
section as prior for the other
parameters, which, as shown in Sect.~6.1, are described well by Gaussian
distributions:

\begin{eqnarray}
\alpha &\sim& \mathcal{N}(-0.12,0.09^2) \\
\sigma_{intr} &\sim& \mathcal{N} (0.18,0.03^2) \\
fact &\sim& \mathcal{N}(-0.22,0.045^2)I(-1,) \quad .
\end{eqnarray}
The stochastic computation, as for
the previous one, is performed by Monte Carlo methods, as explained 
in the Appendix. 

We emphasise that, strictly speaking, the analysis in this
sub--section is using the
data twice: once to derive the posterior probability distribution of $\alpha,\sigma_{intr}$, and
$fact$ (used in Eqs. 19 to 21), and once to infer the $meanZ_{Fe}$. This double use of the
data is conceptually 
wrong and, in general, returns underestimated errors. 
Practically, the information on the $\alpha,\sigma_{intr}$, and
$fact$, derived in Sect. 6.1, is almost independent on
$meanZ_{Fe}$ derived here, and therefore errors are very close to
the correct value.
Readers unsatisfied by this approximation should only rely on the analysis 
presented in Sect.~6.1, which does not make double use of the data.

The right panel of Fig.~4 shows the result of this binning exercise,
after distributing clusters in 5 (solid dots) or 10 (open dots)
redshift bins of  equal cardinality.
The Fe abundance stays constant for a long way, 4 Gyr or so, after which
it decreases, in 
agreement with the rigorous derivation of a 4-6 Gyr e--folding time. 

\section{Discussion}

\subsection{Comparison with the standard analysis}

In the standard analysis, data are 
partly combined (likelihoods are multiplied) 
inside XSPEC (when clusters are grouped by redshift and fitted with tied 
metal abundances) and partly during the metal abundance vs
redshift fitting using $\chi^2$--like techniques.
Neither of the steps allows metal abundances to differ from
cluster to cluster, i.e. neither of them operates correctly with
likelihoods. Therefore it is not a surprise that the standard analysis
does not recover the input value in our simulation of Sect.~2.

In the Bayesian analysis,
we only used XSPEC to estimate
the individual spectrum parameters (metal abundance and temperature)
and we move the combination of the likelihoods 
entirely outside XSPEC, because XSPEC cannot deal with 
the complex task that our
analysis requires. We also avoided the usual $\chi^2$--like techniques, because 
of their inability to handle cases of high complexity. Our approach
has greater
flexibility, and it respects the product axiom of probability
(i.e. likelihood are correctly operated). It recovers the input 
enrichment history, unlike the standard
analysis.

The standard analysis, even if modified, presents other
difficulties: literature best--fit values of real datasets
are rejected by the very same fitted dataset (Sect.~6.2);
literature analyses performed thus far assume independent data, while
real cluster samples have some clusters listed twice (Sect.~5.1). 
We emphasise that the standard two--step analysis has no robust way of 
addressing the
complex features of astronomical data, such as instrument--dependent iron
abundances and $T$--dependent selection effects, or at the very least, none
of previous works has attempted to propagate the
uncertainty related to the instrument--dependent systematics or 
$T$--dependent selection effects in the final result. 
The greater flexibility of the 
Bayesian analysis allows us, instead, to address them.

It should be noted that Anderson et al. (2009) have already hint at
the difficulties related to the commonly used approaches to
data handling.

\subsection{A more complex enrichment history?}

The binned history of metal enrichment
as the sum of step functions (our redshift
binning) offers a flexible approach to the determination of the
cluster metal abundance history, allowing more complex histories than
assumed in Eq.~1. Visual inspection of the points in the right panel of
Fig.~4 suggests that the metal abundance history may be constant up
to $z\sim 0.4 $ (age $\sim9$ Gyr), and that it then decreases at larger redshifts. 
We are, however, unable to firmly establish (or rule out) 
the presence of a break at $z\sim 0.4 $, 
because the additional model parameters introduced to allow this feature are
not completely determined by the data alone. Nevertheless, we note
that this more complex metal abundance history is qualitatively
different from the Balestra et al. (2007) interpretation of their results
based on (a problematic) analysis of a three times smaller sample:
their metal abundance history is said to be flat at $z>0.4$, where our  
analysis suggests a change, and is claimed to be changing at $z<0.4$, where our
measurements return a constant value.

\subsection{A flexible fitting model}

As briefly mentioned in Sect.~5.3, we adopted a log--normal
scatter in metallicity, i.e. a Gaussian scatter in $\log Z_{Fe}$ 
for simplicity, because the
Gaussian function is the simplest solution to breaking the previously
adopted assumption of no scatter. To illustrate the flexibility
of our fitting model and to test the robustness of
our assumption of a Gaussian scatter, we replaced the Gaussian
distribution with a Student's $t$--distribution with ten degrees of freedom
and the unknown scale $s$.
Our fitting model, provided in the Appendix, easily allows this.
We found $s=0.16\pm0.03$ and
identical values and errors for parameters in common between
the old and new model ($Z_{Fe,z=0.2}, \tau,  \alpha, fact$). The standard
deviation of a Student's $t$--distribution with ten degree of freedom
is given by $\sqrt{10/8} s = 0.18\pm0.04$, to be compared with the 
(indistinguishable) result of
the original analysis assuming a Gaussian scatter in $\log Z_{Fe}$,
$\sigma_{intr}=0.18\pm0.03$. The agreement between
the two estimates of the metallicity spread at a given temperature,
18 \%, shows that our analysis is robust to
the precise shape (Gaussian or Student's $t$) of the assumed probability 
distribution of the scatter.

\subsection{Limitations and improvements}

The fitted model can be improved. It is very simple to allow more
complex metal abundance systematics, for example by
introducing a dependency on the metal metal abundance itself, or
allowing more complex enrichment histories.
However, more and better data are needed to constrain
additional parameters.

In our analysis, we did our
best to account for mass--related trends using the temperature $T$. However, this 
does not exhaust all possible selection effects that may be present in available
uncontrolled collections of clusters, such as current samples are (and
our work is not an exception). In particular,
mass--independent selection effects might exist, and
might bias the results. For example, we know that  
cool core clusters have a higher metallicity than non--cool core clusters
in the local universe (e.g. Allen \& Fabian
1998, De Grandi \& Molendi 2001). 
If the same holds at all redshifts and the fraction of
cool core clusters in the sample decreases with increasing redshift (regardless
of the reason for this: either
because we miss them or just because cool core clusters are intrinsically 
less abundant
at higher redshift), then this selection effect produces a decrease in 
the mean metal abundance as function of redshift.

It is unclear whether the above scenario could be tested using core--excised 
metallicities, because the cluster sample composition would not be changed
by flagging the cluster centre. 
Metallicity is an X--ray--measured quantity, because it comes from a
spectral fit to X--ray data. Therefore, the most promising way to collect a
metallicity--unbiased sample is by selecting the objects independently of
their X--ray properties. This requires to avoid X--ray selected
samples, and to consider samples for which the probability of
including a cluster in the sample is
independent of its X--ray properties at a given mass. Optically
selected cluster samples satisfy this request and 
are therefore well suited to studying the cluster enrichment history, as
well other X--ray--related quantities, such as the 
$L_X-$richness (Andreon \& Moretti 2011), the evolution 
of the $L_X-T$ relation (Andreon, Trinchieri \& Pizzolato 2011),
and the fraction of cool--core clusters (Andreon et al. in preparation).

\section{Summary}

This paper aims to derive the metal abundance history of galaxy clusters. Its
determination requires, however, a proper statistical analysis, given that a)
the standard analysis is unable to recover the input metal abundance history
(on simulated data), b) previously found best--fit histories are rejected by
the fitted data, and c) treating upper limits as in previous works returns
steeper  metal abundance histories.

We derived the shape of temperature and metal abundance likelihood 
functions (a log--normal and a normal, respectively). 
For metal abundances, this requires removing the positivity
constraint, which is a default in XSPEC for very low S/N determinations. 
For temperatures, this is the first time to our knowledge that
the non--normal $T$ likelihood has been 
implemented in a regression fitting involving $T$.
To account for possible mass--dependent selection
effects, we allowed metal abundance to depend on $T$, our mass proxy. 
Since we know that clusters have a spread of metal abundances, 
we allowed clusters to have different metal abundances, even at a fixed mass and
redshift. Prompted by
the possible existence of systematics between Chandra- and XMM- derived
metal abundances, we added an additional parameter in our model to account for
systematics. We adopted a history of 
metal abundance that is defined positively at all times (Eq.~1). We fit all the
parameters at the same time in a Bayesian framework, thus allowing each
parameter to ``feel" the effect of the other (to show co--variance). Our
analysis notices the lack of independence of the data used (because some clusters are 
listed twice in the analysed sample) and addresses, for the first time in this field, 
the difficulty of properly using non--independent data  
without sacrificing the information present in the data 
of the duplicate clusters. 

The code for performing the fitting is provided with the paper 
(in the Appendix). It is very flexible and can be easily 
adapted it to its own needs, for example,
to explore other possible modellings. For example, 
changing the metal abundance scatter from log--normal to
a Student's $t$--distribution requires typing less than ten characters.

We analysed metal abundances and temperatures of 130 clusters observed with
Chandra or XMM--Netwon.  Values derived for the 
130 clusters are listed in 
Table 1. About 70\% of them have been re--evaluated in this work.

By fitting the data with our code, 
we found that clusters reach the present--day metal content by a moderate 
increase in metals with time (see lines in Fig.~4),
by a factor 1.5 in the 7 Gyr sampled by the data. 
The metal content we see today in clusters
is, therefore, neither set early in the Universe history, nor 
produced entirely at late times. 
While entropy feedback predates cluster formation (Ponman et al. 1999),
metal abundance feedback does not exhausted at high redshift. 
We also find that the $T$ (mass) dependence is very small, if there is 
any, and that
clusters show an intrinsic $18\pm3$ \% spread in iron abundance. 
As far as we know, this is the first
determination of the intrinsic scatter of cluster metal abundances.
Metal abundances derived with XMM--Newton turns are
$0.78\pm0.045$ times metal abundances derived from Chandra data. 

Finally, we conclude with a word of caution. While our analysis 
accounts for possible mass--dependent
selection effects, we emphasise that 
mass--independent selection effects may exist
and that the studied sample has an unknown 
selection function. If the
selection function were known, it would be very easy to extend our analysis
to account for it.

\begin{acknowledgements} 
I warmly thank Ben Maughan, without whose help this work would not have 
been possible, and the referee, whose comments increased the readability of the 
paper. It is a pleasure to thank Keith Arnaud, Dunja Fabjan, Joseph Hilbe, 
and Paolo Tozzi  for useful
discussions, Fabio Gastaldello for a sharp comment about abundance likelihood
and Tommaso Maccacaro and Ginevra Trinchieri for stimulating comments that 
improved  the paper's presentation.  
I acknowledge the financial contribution from the agreement 
ASI-INAF I/009/10/0.
\end{acknowledgements}

\appendix

\section{JAGS implementation of the models}

The implementation of Eqs. 1 to 11 in JAGS reads as

{\footnotesize
\begin{verbatim}
model {
for (i in 1:length(modeT)) {
 ft[i] <- Abz02*(1-exp(-t[i]/tau))/(1-exp(-11/tau))	 
 Ab[i] ~ dlnorm(log(ft[i]), pow(intrscat,-2))
 Abcor[i] <-Ab[i]*pow(T[i]/5,alpha)*(1+factor*tid[i])
 modeAb[i] ~ dnorm(Abcor[i],pow(sigmaAb[i],-2)) 
 T[i] ~ dunif(1,20)
 modeT[i] ~ dlnorm(log(T[i]),pow(sigmaT[i],-2))
 # for p-value computation 
 Ab.rep[i] ~ dlnorm(log(ft[i]), pow(intrscat,-2))
 Abcor.rep[i] <-Ab.rep[i]*pow(T[i]/5,alpha)*(1+factor*tid[i])	 
 modeAb.rep[i] ~ dnorm(Abcor.rep[i],pow(sigmaAb[i],-2))  
}
Abz02~dunif(0,1)
tau ~ dunif(1,100)
alpha ~dt(0,1,1) 
intrscat ~ dunif(0,1)
factor~dnorm(0.77-1,pow(0.065,-2))I(-1,)
}
\end{verbatim}
}

Comparison of equations and the code shows that
Normal, log-Normal, Student $t$, and Uniform distributions become
{\texttt{dnorm, dlnorm, dt}}, and {\texttt{dunif}}, respectively, and
that we use \texttt{Ab} to denote abundances.
Following BUGS (Spiegelhalter et al. 1996), JAGS  uses 
precisions, $prec = 1/\sigma^2=pow(\sigma,-2)$, in place of variances $\sigma^2$.
The arrow symbol sygnifies ``take the value of". 
X--ray temperatures are zero--pointed to a round number near the temperature
average, 5 keV, for numerical reasons (e.g. speed up convergence) and 
to simplify the interpretation of the found posterior.

To adopt a Student's $t$--distribution with ten degrees of freedom,
{\texttt dt}, it suffices to replace the line starting by {\texttt Ab[i]} 
with

 {\footnotesize
\begin{verbatim}
 lgAb[i] ~ dt(log(ft[i]), pow(intrscat,-2),10)
 Ab[i] <-exp(lgAb[i])    
\end{verbatim}
}
 
To adopt a more complex enrichment history (Sect.~5.2), the line
starting by {\texttt ft[i]} should be edited by putting the
mathematical expression there for the desired enrichment history. 
The user should
also define a prior for each new parameter.

The implementation of the binned model, Eqs. 13 to 21, in JAGS reads

{\footnotesize
\begin{verbatim}
model {
for (i in 1:length(modeT)) {
 modeAb[i] ~ dnorm(Abcor[i],pow(sigmaAb[i],-2)) 
 modeT[i] ~ dlnorm(log(T[i]),pow(sigmaT[i],-2))
 Ab[i] ~ dlnorm(log(meanAb), pow(intrscat,-2))
 Abcor[i] <- Ab[i]*pow(T[i]/5,alpha)*(1+factor*tid[i])
 T[i] ~ dunif(1,20)
}
meanAb ~ dunif(0.,1)
alpha ~dnorm(-0.12,pow(0.09,-2)) 
intrscat ~ dnorm(0.18,pow(0.03,-2))
factor~dnorm(-0.22,pow(0.045,-2))I(-1,)
}
\end{verbatim}
}

\end{document}